\documentclass[12pt]{article}
\usepackage{color}
\usepackage{soul}

\def\fun#1#2{\lower3.6pt\vbox{\baselineskip0pt\lineskip.9pt
\ialign{$\mathsurround=0pt#1\hfil##\hfil$\crcr#2\crcr\sim\crcr}}}

\begin{document} 

\title{\bf{CP violation in D-meson decays
 and the fourth generation}\\ \vspace{1cm}
{\it To Arkady Vainshtein $70^{th}$ birthday}}
\author{ A.N. Rozanov\\
\small{\em CPPM
IN2P3-CNRS-Universite de Mediterranee, Marseille} \\ 
M.~I.~Vysotsky \\
\small{\em ITEP, Moscow}}
\date{}
\maketitle

\begin{abstract}
LHCb collaboration measured CPV at the level of one percent in
the difference of assymetries in
$D^0(\bar D^0)\to \pi^+ \pi^-,  K^+ K^-$ decays. If confirmed on a
larger statistics and final systematics
this would mean New Physics manifestation. The
fourth quark-lepton generation can be responsible for the observed
effect.
\end{abstract}

CP-violating (CPV) asymmetry in $D^0(\bar D^0)\to\pi^+ \pi^-$ decays is defined
as \begin{equation} A_{CP}(\pi^+ \pi^-) \equiv \frac{\Gamma(D^0
\to\pi^+ \pi^-)-\Gamma(\bar D^0 \to\pi^+\pi^-)}{\Gamma(D^0
\to\pi^+ \pi^-)+\Gamma(\bar D^0 \to\pi^+\pi^-)} \label{1}
\end{equation}
and $A_{CP}(K^+ K^-)$ is defined similarly. LHCb collaboration
result looks like  \cite{1}:
\begin{equation}
\Delta A_{CP} = A_{CP}(K^+ K^-) - A_{CP}(\pi^+ \pi^-) = [-0.82 \pm
0.21({\rm stat.}) \pm 0.11 ({\rm syst})]{\rm \%} \;\; 
\footnote{recently CDF confirmed this result getting 
$\Delta A_{CP}=[-0.62 \pm
0.21({\rm stat.}) \pm 0.10 ({\rm syst})]{\rm \%} $  \cite{1111}}
. \label{2}
\end{equation}

In both decays singly Cabibbo suppressed quark tree diagram
dominates, $c\to du\bar d$ and $c\to su\bar s$ correspondingly.
They are proportional to $V_{cd}V_{ud}^* = -(\lambda + iA^2
\lambda^5 \eta)$ and $V_{cs}V_{us}^* = \lambda$, so both are
almost real and have opposite signs  \cite{11}
($\lambda \approx 0.22$,
$A\approx 0.81$, $\eta\approx 0.34$). CPV in both decays is
proportional to the imaginary term in  Cabibbo-Kobayashi-Maskava
(CKM) matrix elements which
occurs in the interference of tree and QCD penguin diagrams. In a
penguin diagram virtual gluon decays to $d\bar d$ pair in case of
$D\to \pi\pi$ and to $s\bar s$ pair in the case of $D\to KK$.
Since the factor which multiplies four quarks operator is
universal in the exact $U$-spin limit we obtain
\begin{equation}
A_{CP}(\pi^+\pi^-) = -A_{CP}(K^+ K^-) \;\; .
\label{eq:3}
\end{equation}

However, in charmed decays $U$-spin symmetry is violated 
considerably and equality (\ref{eq:3}) can be violated 
substantially as well.

Let us consider penguin amplitude in Standard Model. It is
proportional to
\begin{eqnarray}
V_{cd} V_{ud}^* f(m_{d}) + V_{cs} V_{us}^* f(m_s) + V_{cb} V_{ub}^*
f(m_b) = \nonumber \\
= V_{cs} V_{us}^* [f(m_s) - f(m_{d})] + V_{cb}
V_{ub}^*[f(m_b)-f(m_{d})] \;\; , \label{3}
\end{eqnarray}
where due to unitarity of CKM matrix we subtract zero from the
initial expression. For $D$-meson decays
\begin{equation}
f(m_{d})\sim \ln\left(\frac{M_W}{m_c}\right) \; , \;\; f(m_s) \sim
\ln\left(\frac{M_W}{m_c}\right) + O(\frac{m_s^2}{m_c^2}) \; , \;\;
f(m_b)\sim \ln\left(\frac{M_W}{m_b}\right) \;\; . \label{4}
\end{equation}

Taking into account that $V_{cs}V_{us}^*$ is real we get that the
last term in (\ref{3}) dominates in CPV, since $V_{ub}$ has 
large phase; in the difference $f(m_b)-f(m_d)$ big 
log cancels and what remains is close to one:
\begin{equation}
[A_{CP}(\pi^+ \pi^-)]_{SM} = -[A_{CP}(K^+ K^-)]_{SM} \sim |V_{cb}
V_{ub}| \ln\left(\frac{m_b}{m_c}\right) \approx 2\cdot 10^{-4}
\;\; . \label{5}
\end{equation}
This small number makes $\Delta A_{CP} \sim 1$\% highly improbable
in Standard Model.
Naive estimates lead to $\Delta A_{CP} =O(0.05\%-0.1\%)$,
an order of magnitude smaller than the experimental result
\cite{12}.
Nevertheless it  is not excluded that Standard Model explaines 
large $CP$ violation in $D$ decays \cite{33,34}. In order to increase 
$A_{CP}$ in the framework of Standard Model one need to assume very 
big annihilation amplitudes with penguin contraction \cite{13}. 
As such a scenario has very high uncertainty we do not consider it in
the following.

In the case of the fourth generation the second line of (\ref{3})
is substituted by:
\begin{equation}
V_{cs}V_{us}^*[f(m_s)-f(m_{d})]+V_{cb}V_{ub}^*[f(m_b)-
f(m_{d})]+V_{cb'}V_{ub'}^*[f(m_{b'})-f(m_{d})]. \label{6}
\end{equation}
Let us take $m_{b'}=500$ GeV in order to avoid bounds from the
searches of the fourth generation quarks at Tevatron and 
Large Hadron Collider (LHC). For
such heavy $b'$ quark $f(m_{b'})$ is small, $f(m_{b'})\approx
0.15$ (see for example \cite{2}, eqs (A.12), (A.13), 
where explicit dependence of penguin amplitude $F_1\equiv 2f$ 
on the mass of the virtual quark is presented) and can be safely 
neglected in comparison with $f(m_d)$.
Assuming a large phase of the product
$V_{cb'} V_{ub'}^*$ instead of SM estimate (\ref{5}) we obtain:
\begin{equation}
[A_{CP}(\pi^+ \pi^-)]_{4g} = -[A_{CP}(K^+K^-)]_{4g} \sim
|V_{cb'}V_{ub'}^*|\ln\left(\frac{M_W}{m_c}\right) \;\; .
\label{7}
\end{equation}

The value of $|V_{cb'} \cdot V_{ub'}^*|$ is bounded from
above by the measurement of $D^0 - \bar D^0$ oscillation
frequency. Let us suppose that in the case of the fourth
generation this frequency is dominated by the box diagram with
intermediate $b'\bar b'$ quarks. Then according to standard
formula \cite{3}
\begin{equation}
\Delta m_D = \frac{G_F^2 B_D f_D^2}{6\pi^2} m_D m_{b'}^2 \cdot \eta_D
|V_{cb'} V_{ub'}^*|^2 I\left(\frac{m_{b'}^2}{M_W^2}\right) \;\; .
\label{8}
\end{equation}
Substituting $\eta_D = 0.5$, $f_D = 200$ MeV, $B_D =1$,
$I(\frac{m_{b'}^2}{M_W^2}) = 0.25$ and using $\Gamma_D = [4\cdot
10^{-13} {\rm sec}]^{-1}$ we get:
\begin{equation}
x \equiv \frac{\Delta m_D}{\Gamma_D} \approx
0.01\left[\left(\frac{m_{b'}}{1{\rm GeV}}\right)|V_{cb'}\cdot V_{ub'}^*|
\right]^2= 0.01 \;\; , \label{9}
\end{equation}
where the last number is the experimental result. So the product
of CKM matrix elements is bounded by
\begin{equation}
|V_{cb'} \cdot V_{ub'}^*| < 2 \cdot 10^{-3} \;\; , \label{10}
\end{equation}
where the upper value corresponds to the dominance of $(b'\bar
b')$ box in $\Delta m_D$ (see also \cite{333}). 
Comparing (\ref{7}) and (\ref{5}) we see
that the fourth generation can enhance Standard Model result for
$\Delta A_{CP}$ by factor 40 and fit experimental result
(\ref{2}).

From the unitarity bound $|V_{ud}|^2 + |V_{us}|^2 + |V_{ub}|^2 +
|V_{ub'}|^2 = 1$ and numerical values of the first three terms 
from \cite{10} it
follows that $|V_{ub'}| \leq 1.5 \cdot 10^{-2}$ is allowed and
taking $|V_{cb'}| > 0.15$ we can obtain $|V_{cb'}V_{ub'}^*| = 2
\cdot 10^{-3}$.

In conclusion let us note that proposed mechanism can lead to 
CPV in $D^0 - \bar D^0$ mixing at the level of present 
experimental constraints.

   Even if we suppose following \cite{111}  that the product
$|V_{cb'}V_{ub'}^*|$ times  sine of its phase 
is one order of magnitude smaller than what we used, the factor
$\ln (M_W/m_c) \approx 4$ enhancement of $\Delta A_{CP}$
in case of four generations
in comparison with SM result still remains and helps to
explain the experimental number (\ref{2}).

We congratulate our colleague, friend
and penguin discoverer Arkady Vainshtein with his $70^{th}$ birthday.

We are grateful to A.E.Bondar and M.V. Danilov for useful 
discussion and to Giovanni Punzi and Robert Harr for providing
us the CDF result. M.V. is
partially supported by the grant RFBR 11-02-00441 and by the grant
of the Russian Federation goverment 11.G34.31.0047.

\end{document}